\documentclass[twocolumn,preprintnumbers,amsmath,amssymb,prb,floatfix,superscriptaddress]{revtex4-2}

\usepackage[usenames,dvipsnames]{color}
\usepackage{graphicx}
\usepackage{dcolumn}
\usepackage{simplewick}
\usepackage{bm}
\usepackage{amsmath}
\usepackage{braket}
\usepackage[T1]{fontenc}

\usepackage{float}
\usepackage{threeparttable, tablefootnote}
\usepackage{natbib}
\usepackage{multirow}
\usepackage{hyperref}
\hypersetup{
    colorlinks=true,
    linkcolor=blue,
    filecolor=magenta,      
    urlcolor=cyan,
    pdftitle={Overleaf Example},
    pdfpagemode=FullScreen,
    }
\usepackage{booktabs}
\usepackage{siunitx}
\DeclareSIUnit{\angstrom}{\textup{\AA}}
\usepackage{booktabs,multirow,array,tabularx}

\usepackage[utf8]{inputenc}
\usepackage{newunicodechar}
\newunicodechar{⁻}{\textsuperscript{-}}

\begin{document}

\title{Emergence of polar monoclinic phase in heterohalogen substituted
CsGeX$_3$ }
\author{Sourabh Vairat}
\author{Balachandra G. Hegde}
\affiliation{Department of Physics, Rani Channamma University, Belagavi, 591 156, India
}

\author{Brajesh Tiwari}
\author{Ravi Kashikar}
\email{ravikashikar@iitram.ac.in}
\affiliation{Department of Basic Sciences, Institute of Infrastructure, Technology, Research and Management(IITRAM), Ahmedabad, Gujarat 380026, India}

\date{\today}

\begin{abstract}
The occurrence of ferroelectricity in inorganic germanium-based halide perovskites has provided an alternative to oxide counterparts. Using first-principles methods, we have studied CsGeX$_3$ materials with heterohalogen substitution at the X site in a 2:1 configuration. Structurally, such variation alters the octahedral environment more strongly than in pristine materials, giving rise to a polar monoclinic phase at room temperature. The occurrence of the monoclinic phase is also confirmed through the energetics of structures generated by the displacements of atoms in accordance with soft mode eigenvectors of the dynamical matrix along various directions. In the chemically tuned phase, the polarization is along [101] and increases by 10-15\% compared to pristine ones. The electronic structure analysis reveals that spin-splitting energy ranges from 25 to 250 meV in the valence band and from 9 to 80 meV in the conduction band in chemically tuned structures. In addition, these structures exhibit Rashba and persistent spin textures, which are coupled to the polarization direction. The parameters of the symmetry-dependent \textbf{k.p } Hamiltonian provide insights into the strength of spin-splitting and the nature of spin-texture. The semiconducting and spin-polarized nature of CsGeX$_3$ materials makes them strong candidates for Datta-Das spin transistors. 
\end{abstract}
\maketitle

\section{Introduction}
Inorganic halide perovskites of the form (Cs, Rb)BX$_3$ (B = Ge, Sn, Pb; X = Cl, Br, I) are known for light-absorbing materials in optoelectronic devices \cite{tai2019recent,
xiang2021review, han2024inorganic}. Their soft lattice nature, simple synthesis methods, tunable bandgap via chemical substitution, and high defect tolerance offer advantages over oxide perovskites\cite{chiara2021germanium,
guo2024understanding,luo2019b}. Among these, in the bulk phase, the Pb- and Sn-based halide perovskites crystallize in centrosymmetric phases at various temperatures \cite{yang2017spontaneous,
da2015phase}.  However, the Ge-based CsGeX$_3$ and RbGeX$_3$ crystallize in non-centrosymmetric phases, with CsGeX$_3$ in particular exhibiting a polar space group R3m at room temperature and undergo a phase transition to a non-polar cubic phase (Pm$\bar{3}$m) at high temperature\cite{thiele1989kristallstrukturen, yamada1994successive, thiele1987kristallstrukturen, kuang2025theoretical, yang2020assessment}. Experimental demonstration of ferroelectricity in CsGeX$_3$ was shown by Zhang et al, and these materials exhibit a spontaneous polarization of about $12$-$20$ $\mu$C/cm$^2$ \cite{zhang2022ferroelectricity, chen2022ferroelectric}. Using effective Hamiltonian methods, CsGeBr$_3$ was modeled using molecular dynamics and found to host polar-antipolar domains\cite{kashikar2024coexistence}. In the thin film form, the same CsGeBr$_3$ contains stripe, zigzag, and columnar domains due to the depolarizing field\cite{kashikar2024ferroelectricity}. In addition, CsGeX$_3$ [X = Cl, Br, I] exhibit a pyrocoefficient of 1120 to 3800 $\mu$C/$m^2$K, and susceptibility of 300-400 at Curie temperature\cite{kashikar2026dft}. From the electronic structure perspective it has been shown that the CsGeI$_3$ exhibits spin-splitting of 171 meV in CsGeI$_3$ \cite{CGX_Popoola} and charge-spin conversion factor known as Rashba-Eldstein coefficient of 3.45 $\times 10^{10} \, \hbar/\text{A cm} $ and  0.97 $\times 10^{10} \hbar/\text{A cm}$  in the conduction and valence band of CsGeI$_3$, respectively \cite{popoola2025mechanically}.

These extraordinary features, as mentioned above, are proposed and observed in the pristine rhombohedral (R3m) phase of CsGeX$_3$. However, there is plenty of room for other novel properties. In oxide ferroelectric perovskites, variations in chemical composition, stress, and strain have led to the introduction of distinct polar and non-polar phases compared to the pristine state. Among these, the monoclinic and triclinic phases have attracted significant attention due to their enhanced piezoelectric and electromechanical properties in lead-based ferroelectrics (PZT, PMN-PT) \cite{noheda2000stability, noheda2001polarization,noheda2002structure}. The lower-symmetry monoclinic phase enables rotation of the polarization vector, thereby enhancing the piezoelectric response \cite{noheda2000tetragonal}.   In the case of Pb and Sn-based halide perovskites,  alloy engineering at the A, B, or X site leads to bandgap tuning and hence an enhancement in optical absorbance and solar cell efficiency\cite{liu2020b,azizman2024mixed,raoui2021harnessing,
ito2018mixed,
li2025enhanced}.  Also, in CsGeBr$_3$, using effective Hamiltonian methods under mechanical load in the form of uniaxial and biaxial strain has revealed new polar phases, such as tetragonal (P4/mm), orthorhombic (Amm2), and monoclinic (Cm) \cite{townsend2024ferroelectric}.

The CsGeX$_3$ materials are semiconducting and have a bandgap of 1.6 eV to 3.3 eV \cite{zhang2022ferroelectricity}. From an application perspective, these materials are suitable for spintronics applications, as they exhibit spin polarization near the Fermi level in their electronic band structure \cite{CGX_Popoola}. However, the low spin-charge conversion efficiency and Rashba-type spin texture due to C$_{3v}$ symmetry in k-space hinder the devices efficiency \cite{wang2024challenges}. Enhancing spin splitting to reduce thermal effects and inducing persistent spin texture (PST) may provide multifunctional features. The PST is a unique pattern in the k-space\cite{bernevig2006exact}, for which the spin directions are always pointed in a particular direction to any wave vectors\cite{schliemann2017colloquium, kohda2017physics}. Such non-ballistic transport could provide excellent long spin lifetimes, which are highly required for spin-field effect transistor (s-FET)\cite{PhysRevLett.90.146801}. However, such a kind of PST can exist in orthorhombic and monoclinic phases with C$_{2v}$ and C$_s$ point group symmetry \cite{Zunger1,kashikar2023persistent}. The chemical substitution,  at the X site, acts as an uniaxial and biaxial strain, inducing either the orthorhombic or monoclinic phase, as proposed in the effective Hamiltonian study of CsGeBr$_3$ \cite{townsend2024ferroelectric}. Using density functional theory, we aim to (i) unravel and characterize chemically tuned structural transitions in CsGeX$_3$ that emerge from X-site substitution, with an emphasis on symmetry lowering and the stabilization of polar phases; (ii) study the influence of halide substitution on functional properties, including spontaneous ferroelectric polarization, phonon band dispersion and dynamical stability, and the evolution of the electronic band structure; and (iii) propose spin–orbit coupling–induced spin polarization, particularly spin splitting energy, persistent spin texture, and ferroelectric–Rashba co-functionality.

 \section{Computational Details}
Density functional theory (DFT) simulations were performed using the Vienna Ab-initio Simulation package (VASP)\cite{kresse1993ab,kresse1996efficiency,kresse1996efficient,kresse1999ultrasoft}. We employed the Perdew-Burke-Ernzerhof's (PBE)\cite{perdew1997generalized} exchange correlation functional within projected-augmented-wave (PAW) pseudopotential \cite{blochl1994projector}. The plane wave basis was set to 600 eV and a k-mesh of 8$\times$8$\times$8 k-was utilized for Brillouin zone integration. Elemental substitution was performed at the X site in the $2:1$ form, and then structural relaxation was carried out using the conjugate gradient algorithm.  The electronic structure calculations were carried out with and without spin-orbit (SOC) coupling, using an 8$\times$8$\times$8 k-mesh for both pristine and chemically tuned structures. For spin texture calculations, we used a 21$\times$21 k-mesh for a specified plane in reciprocal space as generated by PYPROCAR\cite{herath2020pyprocar}. The phonon dispersions of the cubic unit cell and 2$\times$2$\times$2 supercell were obtained using density functional perturbation theory (DFPT) with the help of PHONONPY \cite{togo2015first, togo2008first}. The polarization calculation was done using the modern theory of polarization as proposed by King-Smith and Vanderbilt\cite{king1993theory}. For symmetry mode analysis, we used FINDSYM \cite{stokes2005findsym} to determine the spacegroup. ISODISTORT\cite{campbell2006isodisplace} from Bilbao crystallography was used to obtain the details of the modes.  All crystal structures were visualized using VESTA \cite{momma2011vesta}. All the calculations, including structural relaxation, polarization, and band structures, were further analyzed using the strongly constrained and appropriately normed revised meta-generalized gradient approximation r$^2$SCAN (Regularized-restored) exchange–correlation functional \cite{furness2020accurate}, to understand the robustness and reliability of the results.

\section{Results and discussion}
\subsection{Structure and phase transition}
\begin{figure*}
        \centering
        \includegraphics[width=0.8\linewidth]{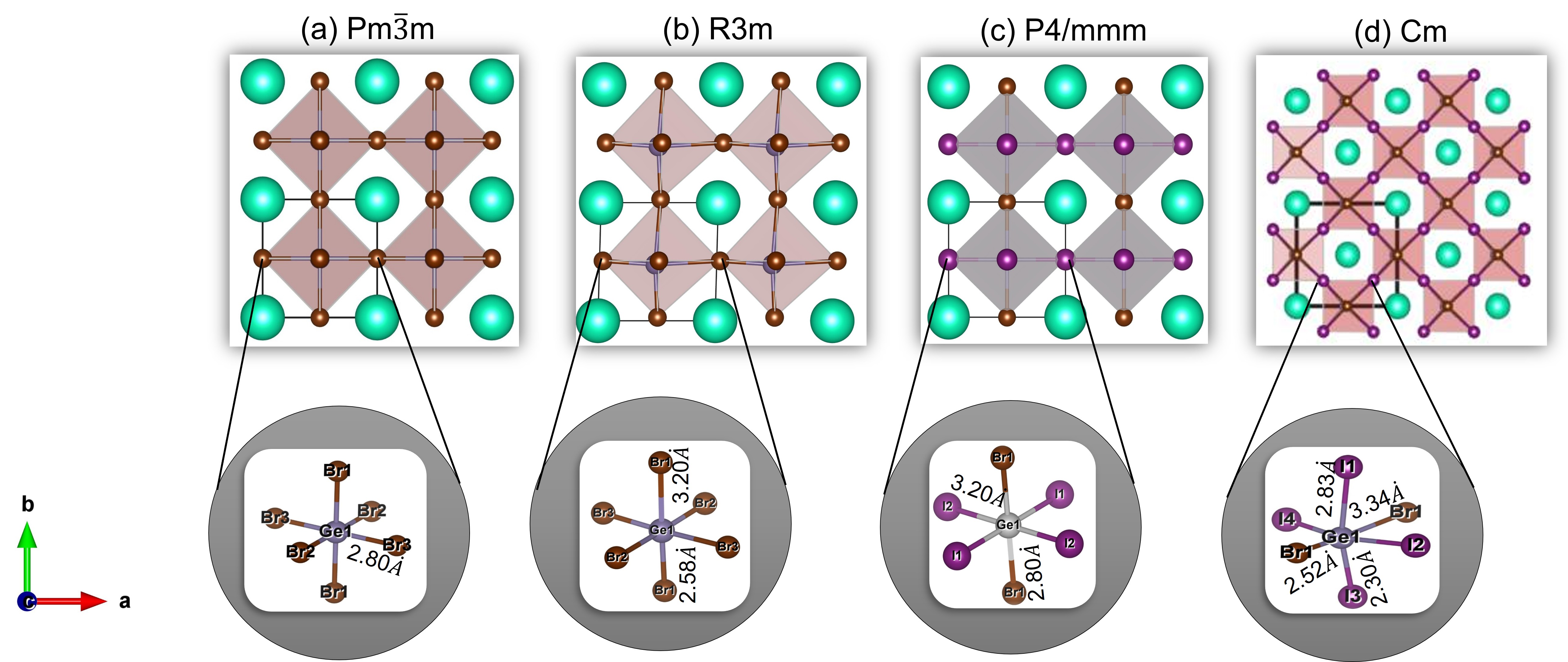}
        \caption{(a)-(b) Structure of CsGeBr$_3$ in paraelectric (Pm$\bar{3}$m) and ferroelectric (R3m) phases.  (c)-(d) Chemically tuned structures of CsGeBrI$_2$ in paraelectric (P4/mmm) and ferroelectric (Cm) phases. }
        \label{fig1}
\end{figure*}

The pristine CsGeX$_3$ (X = Cl, Br, I) perovskites show a single phase transition from the high temperature cubic (Pm$\bar{3}$m) to the low temperature rhombohedral (R$3$m) phase. This phase transition occurs at 428~K for CsGeCl$_3$, 511~K for CsGeBr$_3$, and 550~K for CsGeI$_3$ \cite{thiele1987kristallstrukturen}. The lattice parameters obtained from structural relaxation using the PBE functional are very close to the experimental values, as shown in Table S1 of the supplementary material (SM). The octahedral environment for the cubic and rhombohedral structures is shown in Figs.~\ref{fig1}(a)-(b). In the rhombohedral phase, the Ge atom displaces along the [111] direction with magnitudes of 0.05~$\mathrm{\mathring{A}}$ (CsGeCl$_3$), 0.1~$\mathrm{\mathring{A}}$ (CsGeBr$_3$), and 0.17~$\mathrm{\mathring{A}}$ (CsGeI$_3$), leading to asymmetric compressed and elongated bonds (Fig.~\ref{fig1}(b)), making CsGeX$_3$ family polar inorganic perovskites. Experimentally, ferroelectricity in CsGeX$_3$ has been studied by Zang et al., and reported that CsGeBr$_3$ exhibits spontaneous polarization of $12$-$15$ $\mu$C/cm$^2$, and in CsGeI$_3$ of $20$ $\mu$C/cm$^2$, respectively\cite{zhang2022ferroelectricity,chen2022ferroelectric}. From the density functional theory using the PBE exchange-correlation functional, the reported spontaneous polarization values are in Refs. \onlinecite{kashikar2026dft, kashikar2024coexistence, CGX_Popoola}. Our simulated values agree very well with these previously reported results. The theoretical values of spontaneous polarization decrease as we move from CsGeCl$_3$ to CsGeBr$_3$ to CsGeI$_3$. However, experimentally, the trend is opposite \cite{zhang2022ferroelectricity}. The phonon calculations in the high-symmetry phase reveal the existence of soft modes at M, $\Gamma$, and X, and among them the largest negative frequency, which is triply degenerate,  occurs at $\Gamma$, corresponding to the $\Gamma_{4}^{-}$ irreducible representation.  The opposite trend of polarization and Curie temperature may be due to competing instabilities at $\Gamma$ and X points\cite{kashikar2026dft}. The eigenvector corresponding to the largest negative frequency indicates an opposite directional displacement of the Ge and Br atoms, leading to the polar R3m phase.

The splitting of the triply $\Gamma_{4}^{-}$ mode into a nondegenerate $\Gamma_{3}^{-}$ mode and a doubly degenerate $\Gamma_{5}^{-}$ mode is necessary to achieve different polar phases. The condensation of these modes drives the system into either a polar orthorhombic phase (associated with $\Gamma_{3}^{-}$) or a polar monoclinic phase (associated with $\Gamma_{5}^{-}$). This is possible by creating an anisotropic octahedral environment. One way to create such anisotropy is through the application of an uniaxial or biaxial strain. Using the effective Hamiltonian technique, Townsend et al., reported that mechanical load in the form of uniaxial or biaxial stress induces a monoclinic phase during the structural phase transition, in addition to the enhancement of spontaneous polarization in CsGeBr$_3$ \cite{townsend2024ferroelectric}. Chemical substitution of hetero-halogens at the X site in pristine CsGeX$_3$ effectively acts as an internal uniaxial or biaxial chemical strain. The size and electronegativity mismatch among the substituted halide ions modifies the local bonding environment and lattice parameters, thereby distorting the GeX$_6$ octahedra. These effects break the parent crystal symmetry and can drive structural instabilities, ultimately inducing symmetry-lowering structural phase transitions and stabilizing new crystallographic phases.

\begin{figure*}
    \centering    \includegraphics[width=0.8\linewidth]{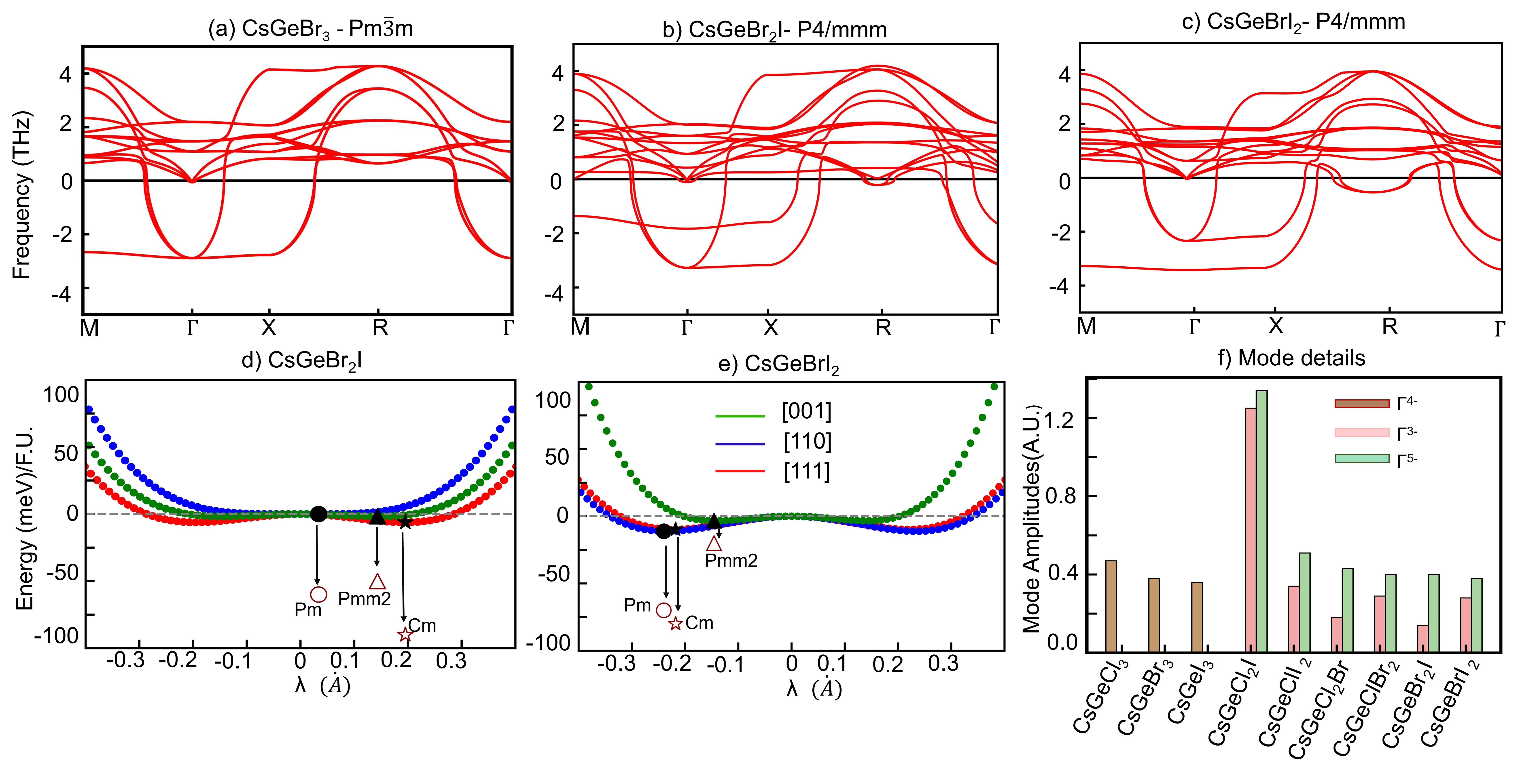}
    \caption{(a)-(c) Phonon spectrum for centrosymmetric Pm$\bar{3}$m-CsGeBr$_3$, P4/mmm-CsGeBr$_2$I and CsGeBrI$_2$. (d)-(e) Energy variation along distortion for [001], [110], and [111] displacements for non-polar P4/mmm-CsGeBr$_2$I and CsGeBrI$_2$. (f) Amplitude of structural distortions in different modes for pristine and chemically tuned compounds.}
    \label{fig2}
\end{figure*}

\begin{figure}
    \centering
    \includegraphics[width=1\linewidth]{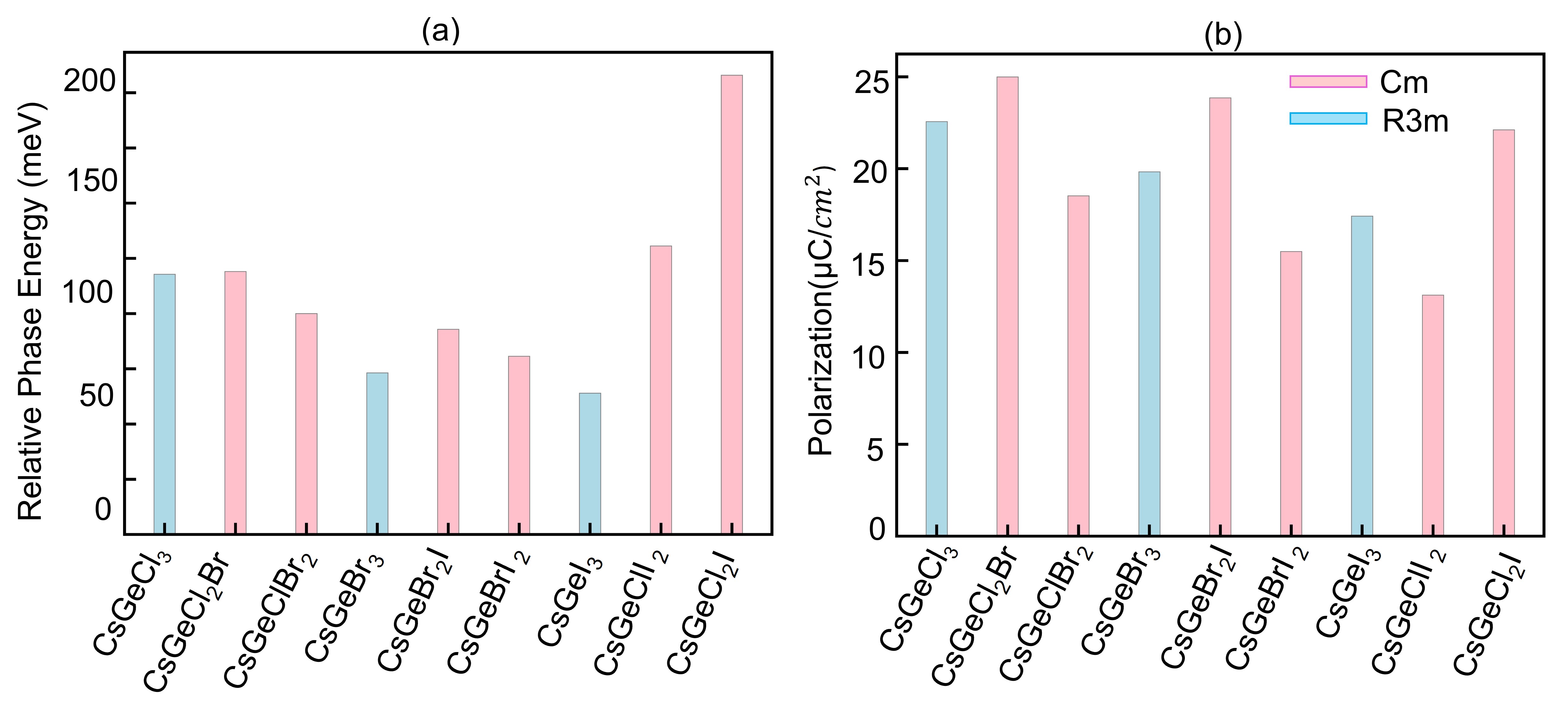}
    \caption{a) Relative Phase Energy for pristine and various chemical compositions. (b) Spontaneous polarization of pristine and various chemical compositions for R3m and Cm phases, respectively.}
    \label{fig3}
\end{figure}
 We carried out the structural relaxation on cubic CsGeX$_3$, with halogen; Cl, Br, or I in a 2:1 form at the X site, and observed a centrosymmetric P$4$/mmm phase. The hetero-halogen substitution introduces an anisotropic bonding environment within the octahedron, as shown in Fig.~\ref{fig1}(c). The phonon spectrum of chemically tuned structures in the centrosymmetric P$4$/mmm phase is shown in Fig.~\ref{fig2}(b)-(c). Similar to the pristine phase, the chemically tuned structures also exhibit soft modes at the $\Gamma$, M, and X points. However, the triply degenerate mode with negative frequency at $\Gamma$ has now been lifted. For example, in one of the chemically tuned structures, CsGeBr$_2$I, the largest negative frequency is doubly degenerate, whereas in CsGeBrI$_2$ it is nondegenerate.

 To study the possible phase transition that the chemically tuned centrosymmetric P$4$/mmm phase may undergo due to lowering of the temperature, we displaced the various atoms along [001], [110], and [111] directions in accordance with the eigenvector of the largest negative eigenvalue from the dynamical matrix. The energy profile along these directions for CsGeBr$_2$I and CsGeBrI$_2$ is shown in Fig.~\ref{fig2}(d)-(e). Among these, the structures displaced along [111] are lower in energy in CsGeBr$_2$I and in CsGeBrI$_2$. The volume-cell relaxation of the structure corresponding to the minimum-energy for [111] displacement has the polar Cm space group, [110] displacement has the Pm space group, and [001] has the Pmm2 space group, respectively.  Among these, the monoclinic Cm is lower in energy by 90 meV for CsGeBr$_2$I and 80 meV in CsGeBrI$_2$ with respect to the P4/mmm phase, respectively.  This confirms that the monoclinic Cm phase becomes the most stable configuration relative to the P4/mmm phase. The polar monoclinic (Cm) phase is also confirmed through the full structural relaxation in pristine rhombohedral CsGeX$_3$ through hetero-halogen substitution at the X site in 2:1 form.  Anisotropic displacement of the Ge atom and its bonding environment within the octahedron leads to the monoclinic Cm phase. For example, in the case of CsGeBr$_2$I, the Ge atom displaces 0.05$\mathrm{\mathring{A}}$ along the x-axis, and 0.13$\mathrm{\mathring{A}}$ along the y and z axes. For CsGeBrI$_2$, the displacement is 0.09$\mathrm{\mathring{A}}$ along the x-axis, while along the y and z-axes it is 0.11$\mathrm{\mathring{A}}$.  The Cm phase is shown in Fig.~\ref{fig1}(d).
 
To quantify the ferroelectric ordering in the monoclinic phase, we have estimated the spontaneous polarization along the distortion path connecting the paraelectric P4/mmm phase. The relative phase energy among the pristine materials decreases from CsGeCl$_3$ to CsGeI$_3$, falling between 117 and 64 meV. In the case of chemically tuned structures, the relative phase energy falls between 217 and 80 meV, corresponding to CsGeCl$_2$I and CsGeBrI$_2$, respectively, as shown in Fig.~\ref {fig3}(a), confirming the tunability of transition temperatures and coercive fields. The spontaneous polarization, which rotates from the diagonal [111] to the in-plane [101] directions due to the anisotropic environment in the octahedron, has a range of 13.15 $\mu$C/cm$^2$ to 25 $\mu$C/cm$^2$, corresponding to CsGeClI$_2$ and CsGeCl$_2$Br for the PBE functional, as represented in the bar chart Fig.~\ref{fig3}(b). Unlike in oxide perovskites, where a high relative phase energy corresponds to a higher critical temperature and coercive field, we observed that in chemically tuned compounds, there is no such correspondence between the relative phase energy and polarization. Therefore, Ge-based halide perovskites offer the opportunity to tune the polarization and coercive field. 

 The distortion mode analysis for chemically tuned structures in polar phases compared with their paraelectric phase, consists of strain modes, $\Gamma_{1}^{+}$, $\Gamma_{3}^{+}$, $\Gamma_{5}^{+}$, and ferroelectric modes such as $\Gamma_{4}^{-}$, $\Gamma_{5}^{-}$, and $\Gamma_{3}^{-}$. The ferroelectric amplitude of these modes for pristine and chemically tuned material analysis is shown in Fig.~\ref{fig2}(f). Along with these distortion modes, polarization enhancement arises from a significant change in the Born effective charges induced by an anisotropic octahedral environment. The Born effective charges of all the chemically tuned structures are listed in Table S4 of the SM.  To evaluate the robustness of our results, we reevaluated the polarization values of the chemically tuned structure using the meta-GGA r$^2$SCAN exchange–correlation functional. This approach yields slightly higher polarization values compared to the PBE functional, while preserving the same overall variation trend.

\subsection{Electronic properties}
\begin{figure}
    \centering
    \includegraphics[width=1\linewidth]{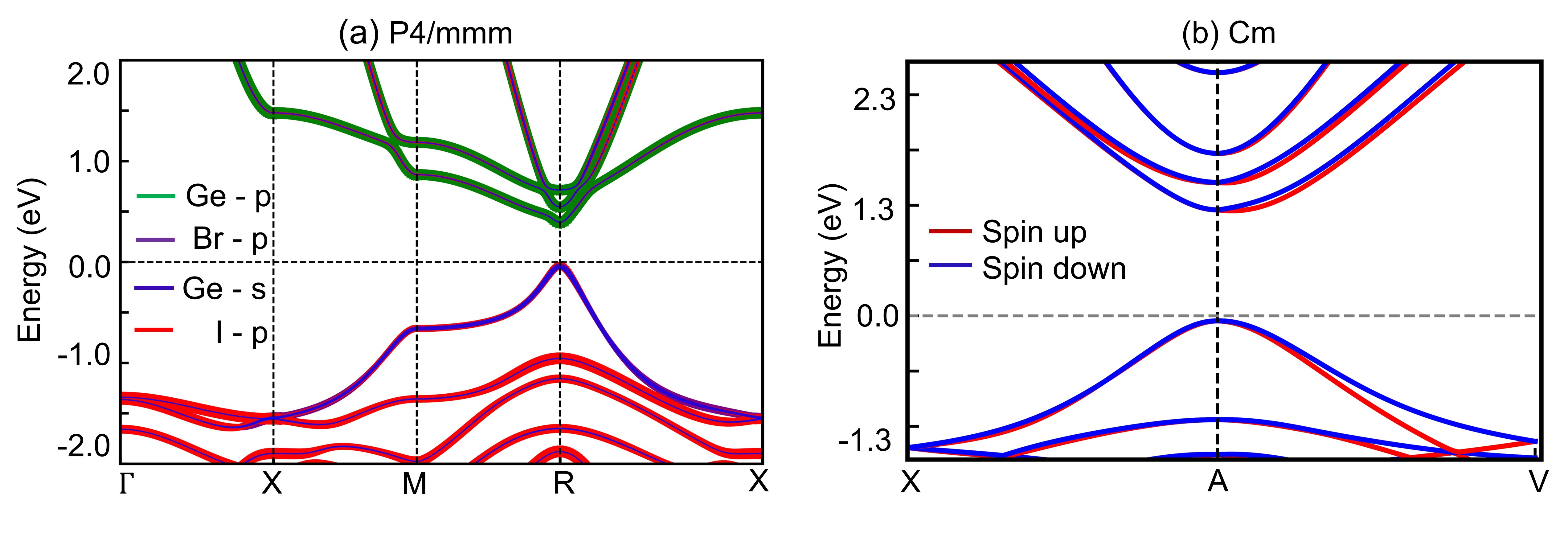}
    \caption{a) Orbital resolved electronic bands structure for chemically tuned structures of CsGeBrI$_2$ in paraelectric (P4/mmm), b) and ferroelectric (Cm) phase.}
    \label{fig4}
\end{figure}

 Having understood the crystal structure and its phase transition in chemically tuned CsGeX$_3$ structures, we now analyze the electronic band structures of these materials. The details of the electronic band structure of pristine CsGeX$_3$ in polar and nonpolar phases have been discussed in detail in the Ref. \onlinecite{CGX_Popoola}. In pristine paraelectric and ferroelectric phases, the conduction bands (CB) are from Ge-p states. In the cubic phase, we observe doubly degenerate p$_{3/2}$ bands and single-degenerate p$_{1/2}$ bands, and they split due to spin-orbit coupling (SOC). In the polar phase, due to structural asymmetry, there is further breakdown of the p$_{3/2}$ band into nondegenerate bands. On the other side, the valence bands (VB) emerging from X-p and Ge-s states. The contribution of atoms and their orbitals near the valence band maximum (VBM) and conduction band minimum (CBM) is listed in the Table S3 (SM). In the chemically tuned structures, P4/mmm is the high-temperature phase, and Cm is the room-temperature polar phase. In both cases, we have non-degenerate CB's arising from both SOC and structural asymmetry. In the P4/mmm phase, the VBM and CBM occur at (0.5, 0.5, 0.5), and in the polar phase, for all chemically tunable structures, the VBM and CBM are at (0, 0, 0.5), which is shown in the Fig.~\ref{fig4} (a)-(b). The orbital contribution near the VBM and CBM for both paraelectric and ferroelectric phases is listed in Table S3 of the SM. The bandgap obtained from the PBE exchange-correlation functional along the paraelectric-to-ferroelectric phase transition exhibits linear behavior. However, the chemically tunability at the X site exhibits nonlinear behavior in its bandgap variation.  The PBE exchange-correlation functional is known to underestimate the bandgap. Usage of the meta-GGA r$^2$SCAN exchange-correlation functional provided a slight increase in value relative to the PBE functional. However, the nature of the variation remains the same along the phase transition path and with halogen substitution. The virtue of ferroelectric ordering has direct implications for the band structure, as the symmetry-breaking field acts as a perturbation that lifts the spin-up and spin-down degenerate bands in the k-space other than the time-reversal invariant momenta.

\begin{figure*}
    \centering
    \includegraphics[width=1\linewidth]{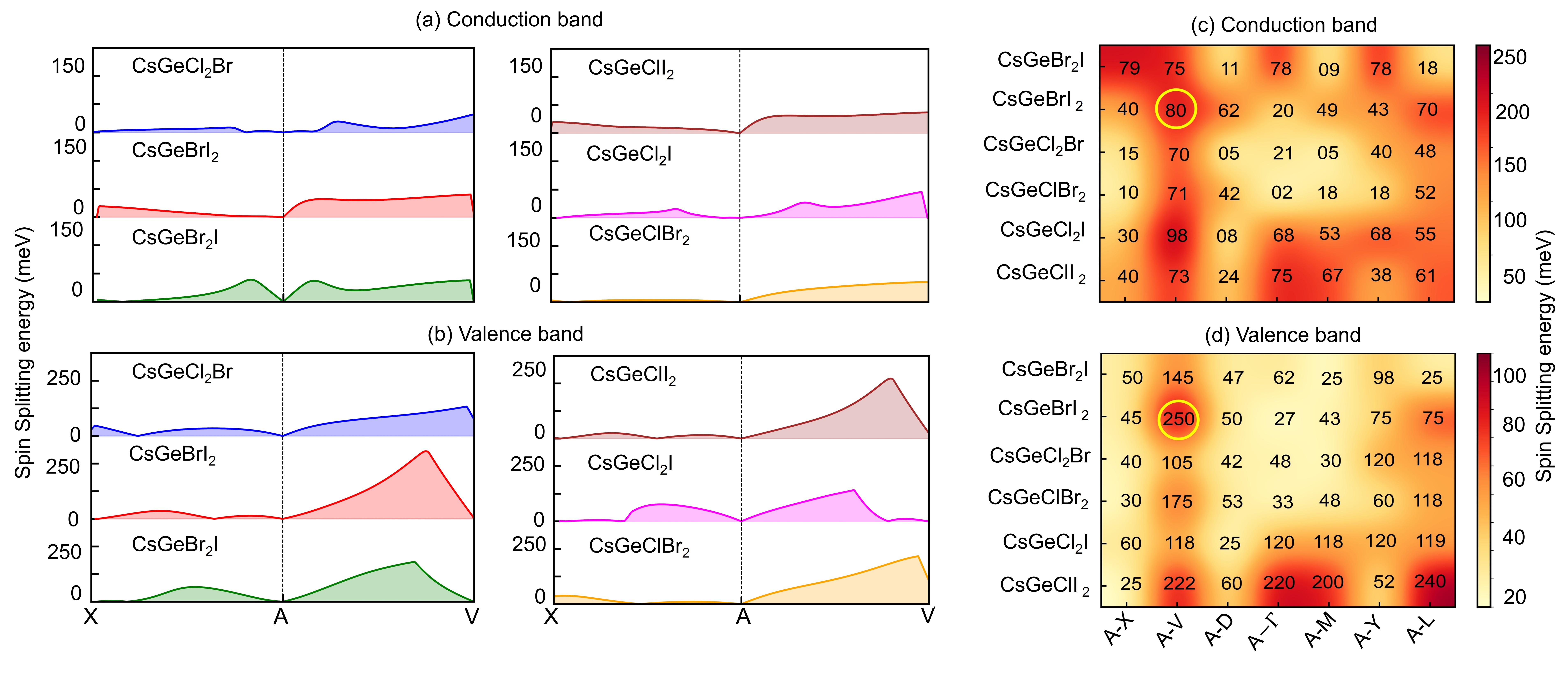}
    \caption{(a)-(b) Spin splitting energy along high symmetry path in the Brillouin zone for conduction band and valence band for all chemically tuned structures. (c)-(d) Anisotropic spin splitting energy for various chemically tuned structures in the conduction and valence bands, respectively.}
    \label{fig5}
\end{figure*}

 The combination of inversion symmetry breaking and spin-orbit coupling breaks the energy degeneracy of the spin-up and spin-down bands. To estimate the spin-splitting energy in the chemically tuned ferroelectric phase, we carried out band structure analysis along various high symmetry k-paths passing through the A point where the band extremum occurs, and it is shown in Fig.~\ref{fig5}(a)-(b). Previously, Popoola et al., have investigated the spin-polarized band structure of pristine compounds and showed that the largest spin-splitting of 171 meV occurs in the valence band along M$\longrightarrow$K in CsGeI$_3$\cite{CGX_Popoola}.  In chemically tuned structures, the polar CsGeBrI$_2$ material in Cm phase exhibits the largest spin-splitting of ~80 meV and ~250 meV along A$\longrightarrow$V in the conduction and valence band, respectively, as presented in Fig.~\ref{fig5}(a)-(b). For other compounds, along various k-paths, the spin-splitting energy results are summarized in the heat maps in Fig.~\ref{fig5}(c)-(d), which indicate that the spin-splitting in these materials is highly anisotropic. The spin-splitting behavior remains nearly identical for PBE and r$^2$SCAN functionals, indicating that the underlying spin–orbit interaction strength is not significantly affected.
 To gain more understanding of various couplings that lead to spin-splitting, we have developed a symmetry-based \textbf{k.p} model. The A point exhibits the C$_s$ point group symmetry, which consists of the mirror plane  M$_y$ and time-reversal symmetry $\Theta$. The \textbf{k.p} Hamiltonian for C$_s$ point group symmetry is presented in Ref.\cite{kashikar2021chemically,kashikar2023persistent}, and it is given as 
\begin{equation}
H =
\frac{\hbar^{2}}{2}
\left(
\frac{k_x^{2}}{m_x}
+
\frac{k_y^{2}}{m_y}
+
\frac{k_z^{2}}{m_z}
\right)
+\alpha k_x \sigma_y
+k_y\left(\beta \sigma_x+\gamma \sigma_z\right)
+k_z \delta \sigma_y
\tag{1}
\end{equation}
where k$_i$ is the cartesian component of the k-vector in the Brillouin zone, m$_x$, m$_y$ and m$_z$ are the effective masses along different directions. $\sigma_i$ are the Pauli matrices, $\alpha$, $\beta$, $\gamma$, and $\delta$ are the spin-momentum locking coupling parameters. To understand the spin-splitting along the k$_x$-k$_y$ plane, the matrix form of the Hamiltonian is written as

\begin{equation}
H =
\begin{pmatrix}
\dfrac{\hbar^{2}}{2}
\left(
\dfrac{k_x^{2}}{m_x}
+
\dfrac{k_y^{2}}{m_y}
\right)
+ \gamma k_y
&
\beta k_y - i \alpha k_x
\\[6pt]
\beta k_y + i \alpha k_x
&
\dfrac{\hbar^{2}}{2}
\left(
\dfrac{k_x^{2}}{m_x}
+
\dfrac{k_y^{2}}{m_y}
\right)
- \gamma k_y
\end{pmatrix}
\tag{2}
\end{equation}
The parameter $\alpha$ can be obtained by fitting the energy bands along D(0.5, 0, 0.5)-A(0, 0, 0.5)-D(0.5, 0, 0.5). Similarly, $\beta$, $\gamma$ and $\delta$ are obtained through the bands along L(0, 0.5, 0.5)-A(0, 0, 0.5)-L(0, 0.5, 0.5), and $\Gamma$(0, 0, 0)-A(0, 0, 0.5)-$\Gamma$(0, 0, 0) respectively. These parameters for CB and VB for CsGeBrI$_2$ and CsGeBr$_2$I are shown in Table I and the fitted bands are shown in Fig. S1 in SM.

\begin{table}[ht]
    \centering
    \caption{Anisotropic hopping parameters $t_i = \frac{\hbar^2}{2m_i}$ where (t$_i$ = $t_x, t_y, t_z$) in $\mathrm{eV \cdot \mathring{A}^{-2}}$ and spin-splitting coefficients ($\alpha$--$\delta$) in $\mathrm{eV \cdot \mathring{A}}$.}
    \label{TableI}
    \begin{tabular}{ll ccc cccc}
        \toprule
        Compound & Band & $t_x$ & $t_y$ & $t_z$ & $\alpha$ & $\beta$ & $\gamma$ & $\delta$ \\
        \midrule
        CsGeBrI$_2$ & CB & 30 & 11.8 & 3.72 & 0.28 & -0.15 & -0.7 & 0.095 \\
                    & VB & -28.8 & -23.7 & -7.34 & 0.075 & -0.075 & 0.15 & 0.1 \\ 
        \addlinespace
        CsGeBr$_2$I & CB & 5.6 & 3.8 & 42.5 & 0.06 & -0.07 & -0.07 & 0.40 \\
                    & VB & -12.0 & -10.5 & -41.7 & 0.085 & -0.03 & 0.14 & 0.1 \\ 
        \bottomrule
    \end{tabular}
\end{table}

\begin{figure*}
     \centering
     \includegraphics[width=1\linewidth]{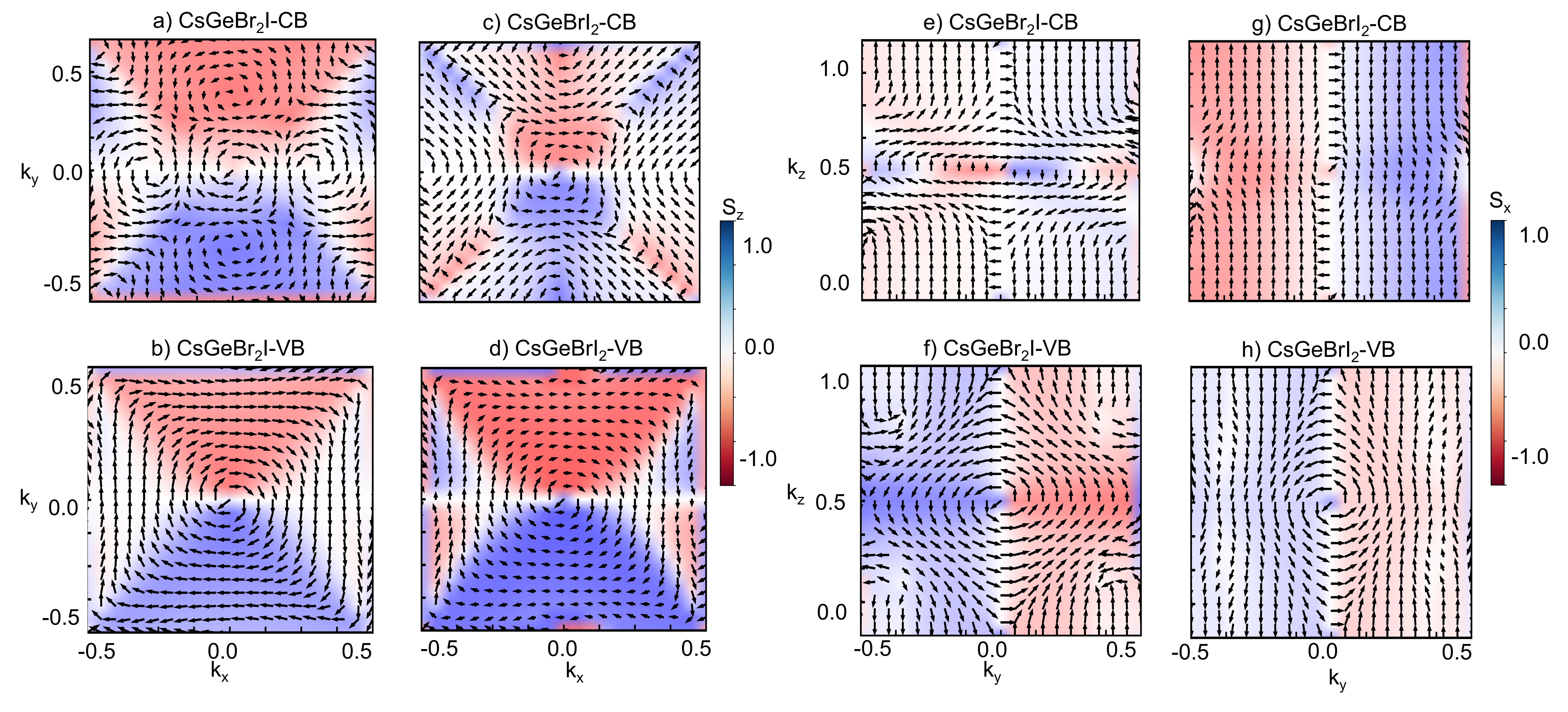}
     \caption{(a)-(b) and (c)-(d) DFT Spin textures for CsGeBr$_2$I and CsGeBrI$_2$ in CB and VB respectively along k$_x$-k$_y$ plane. Similarly, (e)-(f) and (g)-(h) for CsGeBr$_2$I and CsGeBrI$_2$ in CB and VB along k$_y$-k$_z$ plane, respectively.}
     \label{fig6}
 \end{figure*}

 Having understood the spin splitting in k-space, we now analyze the spin texture in different planes of the Brillouin zone. In the case of pristine CsGeX$_3$, it is bound to the expected Rashba or Dresselhaus type of spin-texture in various planes due to C$_{3v}$ point group symmetry. In the chemically tuned monoclinic phase with C$_s$ point group symmetry, there is no restriction on the type of spin texture that it exhibits\cite{mera2021different}.  For C$_s$ point group symmetry, spin textures are expected to be along (k$_x$-k$_y$) and (k$_y$-k$_z$) planes, which are perpendicular to the M$_y$ mirror plane. Fig.~\ref{fig6} represents the spin textures from first-principles methods in the (k$_x$-k$_y$) and (k$_y$-k$_z$) planes near the CBM and VBM for two representative compounds CsGeBr$_2$I and CsGeBrI$_2$. On the (k$_x$-k$_y$) plane, we observed different types of spin textures in CB and VB.  At the lowest CB, we observed a Rashba-type spin texture near the CBM and quasi-persistent type spin texture away from it, where a band intersection occurs.  In the topmost valence band, we have a Rashba-type splitting near the band extremum. In the k$_y$-k$_z$ plane, we observed nearly persistent spin texture in the lowermost CB of CsGeBr$_2$I and CsGeBrI$_2$. In the case of VB, both representative compounds exhibit Rashba-type spin texture in k$_y$-k$_z$ plane. The spin textures for other chemically tuned structures are shown in Fig. S3 of SM. We observed PST in the lowermost CB for CsGeClBr$_2$ and CsGeClI$_2$ structures.

The occurrence of such a variety of spin textures can be understood from the parameters of the \textbf{k.p} model presented in Table~\ref{TableI}. The Rashba-type spin texture on k$_x$-k$_y$ around the VBM and CBM is due to comparable values of |$\alpha$| and |$\beta$|, which yield nearly the same expectation values of $\sigma_y$ and $\sigma_x$. The opposite sign ensures the Rashba type. Similar findings apply to Rashba-type spin texture around the VBM of k$_y$-k$_z$. Comparable values of $\delta$ and $\gamma$ yield nearly the same $\sigma_y$ and $\sigma_z$ components.   The PST around CBM on the k$_y$-k$_z$ plane is due to the dominant value of $\gamma$ over $\delta$ for CsGeBrI$_2$, which offers a dominant $\sigma_z$ over $\sigma_y$. Similarly, for CsGeBr$_2$I, the PST along the y direction stems from a large $\delta$ value over $\gamma$. 
 
\section{Conclusions}

In summary, using a first-principles computational methodology, we investigated the emergence of the polar monoclinic phase in CsGeX$_3$ via heterohalogen substitution at the X site. The substitution of heterohalogens at high temperature cubic phase results in breaking of O$_h$ to D$_{4h}$ symmetry and thereby lifts the phonon soft mode degeneracy. The displacement of atoms along various directions, according to the soft-mode eigenvector, results in a monoclinic phase. We also observed that the heterohalogens at the X site in the pristine rhombohedral CsGeX$_3$ break local octahedral symmetry and induce the Cm phase due to unaxial and biaxial strain within octahedra. The monoclinic Cm phase enhances polarization and has a significant effect on the electronic band structure. The spin-splitting energy, which stems from the polar nature of the materials reaches to 250 meV in VB and 80 meV in CB for CsGeBrI$_2$.  The chemically tuned compounds, which exhibit ferroelectricity with C$_s$ point-group symmetry, host various types of spin textures that are reversible with the polarization. The \textbf{k.p} model developed for this point group symmetry allows us to explore the occurrence of such large spin-splitting and persistent spin texture on k$_y$-k$_z$ plane. Thus, chemical tunability not only enables polarization control but also enables anisotropic spin splitting and spin-texture tuning, which are hindered in pristine compounds.

\section*{Acknowledgement}   
RK acknowledges the support of the Centre of Excellence AI and ML, IITRAM, Ahmedabad.

\onecolumngrid

\newpage

\begin{center}
   \textbf{\Large Supplementary Material}
\end{center}

  \renewcommand{\thefigure}{S\arabic{figure}}
\setcounter{figure}{0}

  \renewcommand{\thetable}{S\arabic{table}}
\setcounter{table}{0}

\begin{table}[H]
  \centering
  \caption{DFT based structural parameters of pristine and chemically tuned structures for PBE functional.}
  \label{tabS1}
  \small
  \begin{tabularx}{\textwidth}{|l|*{3}{>{\centering\arraybackslash}X|}|*{3}{>{\centering\arraybackslash}X|}}
    \hline
    \textbf{Compound} & \multicolumn{3}{c|}{\textbf{Low temperature}} & \multicolumn{3}{c|}{\textbf{High temperature}} \\
\hline
 & lattice parameter (~\AA) & Space group & Bandgap (eV) & lattice parameter (~\AA) & Space group &  Bandgap (eV) \\
\hline
CsGeCl$_3$ & 5.53 &  R$3$m & 2.01 & 5.35
 & Pm$\bar{3}$m & 1.10\\
CsGeCl$_2$Br & 8.71, 8.50, 5.86 & Cm & 1.57 & 5.60, 5.35, 5.35
 & P$4$/mmm & 0.73 \\
CsGeClBr$_2$ &8.21, 8.02, 5.58
 & Cm & 1.45 & 5.61, 5.33, 5.61  & P$4$/mmm & 0.65 \\
CsGeBr$_3$ & 5.75 & R$3$m & 1.32 & 5.60 & Pm$\bar{3}$m & 0.75 \\
CsGeBr$_2$I & 8.35,	8.15,	6.02
 & Cm & 1.07 & 5.97,	5.63,	5.63
 & P$4$/mmm & 0.50 \\
CsGeBrI$_2$ & 8.71,	8.49,	5.86
 & Cm & 0.99 & 6.00,	5.60,	6.00
 & P$4$/mmm & 0.43 \\
CsGeI$_3$ & 6.11
 & R$3$m & 0.93 & 5.99
 & Pm$\bar{3}$m & 0.64 \\
 CsGeClI$_2$ & 8.72,	8.51,	5.70
 & Cm & 1.08 & 6.04,	5.32,	6.04
 & P$4$/mmm & 0.46 \\
CsGeCl$_2$I & 8.16,	7.96,	6.13
 & Cm & 1.38& 5.96,	5.40,	5.40 & P$4$/mmm & 0.68 \\

    \hline
\end{tabularx}
\end{table}

\begin{table}[H]
  \centering
  \caption{DFT based structural parameters of selected  chemically tuned structures from r$^2$SCAN functional.}
  \label{tabS2}
  \small
  \begin{tabularx}{\textwidth}{|l|*{3}{>{\centering\arraybackslash}X|}|*{3}{>{\centering\arraybackslash}X|}}
    \hline
    \textbf{Compound} & \multicolumn{3}{c|}{\textbf{Low temperature}} & \multicolumn{3}{c|}{\textbf{High temperature}}  \\
\hline
 & lattice parameter (~\AA) & Space group & Bandgap (eV) & lattice parameter (~\AA) & Space group &  Bandgap (eV) \\
\hline

CsGeCl$_2$Br & 7.80, 7.71, 5.64 & Cm & 1.73 & 5.60, 5.34, 5.34
 & P$4$/mmm & 1.07  \\

CsGeBr$_2$I & 8.23,	8.04,	5.96
 & Cm & 1.19 & 5.58,	5.58,	5.91
 & P$4$/mmm & 0.54 \\
CsGeBrI$_2$ & 8.65, 8.41, 5.76 
 & Cm &  1.08  & 5.56,	5.95,	5.56
 & P$4$/mmm & 0.54 \\

    \hline
\end{tabularx}
\end{table}

\begin{table}[H]
  \centering
  \caption{Orbital contribution in pristine and in chemically tuned compounds near conduction band minima (CBM) and valence band maxima (VBM).}
  \label{tabS3}
  \small
  \begin{tabularx}{\textwidth}{|l|*{2}{>{\centering\arraybackslash}X|} *{2}{>{\centering\arraybackslash}X|}}
    \hline
    \textbf{Compound} & \multicolumn{2}{c|}{\textbf{Non-Polar}} & \multicolumn{2}{c|}{\textbf{Polar}} \\ \hline
    & CBM & VBM & CBM & VBM \\ \hline
    CsGeBr$_3$   & Ge-p ~ 83\%,  (Ge, Br)-(s,p) ~ 17\%, & Ge-p ~ 38\%, Br-p ~ 62\%  & Ge-p ~70\%, (Ge, Br)-(s,p) ~ 30\% & Br-p ~55\% , Ge-(s,p) ~45\%  \\ \hline
    CsGeBrI$_2$  & Ge-p ~ 83\%, (Br, I)-(s,p) ~17\% & Ge-s ~36\%, (Br, I)-p ~64\%  & Ge-p ~75\%, (Br, I)-(s,p) ~25\%  & Ge-s ~37\%, (Br, I)-(s,p) ~63\% \\ \hline
  \end{tabularx}
\end{table}

\begin{table}[H]
\centering
\caption{Born Effective Charges ($Z^*$) for pristine and chemically tuned compounds}
\vspace{5pt}
\scriptsize
\setlength{\tabcolsep}{3pt}
\renewcommand{\arraystretch}{1.3}

\begin{tabularx}{\textwidth}{|>{\raggedright\arraybackslash}p{1.8cm}|*{9}{>{\centering\arraybackslash}X|}}
\hline
\textbf{Compound} 
& CsGeCl$_3$ 
& CsGeCl$_2$Br 
& CsGeClBr$_2$ 
& CsGeBr$_3$ 
& CsGeBr$_2$I 
& CsGeBrI$_2$ 
& CsGeI$_3$ 
& CsGeClI$_2$ 
& CsGeCl$_2$I \\
\hline

\textbf{$Z^*$ (e)} 
& 5.10
& 5.06 
& 5.83 
& 6.16 
& 7.38 
& 7.68 
& 8.48 
& 5.38 
& 7.32 \\
\hline
\end{tabularx}

\end{table}

\begin{figure}
    \centering
    \includegraphics[width=0.8\linewidth]{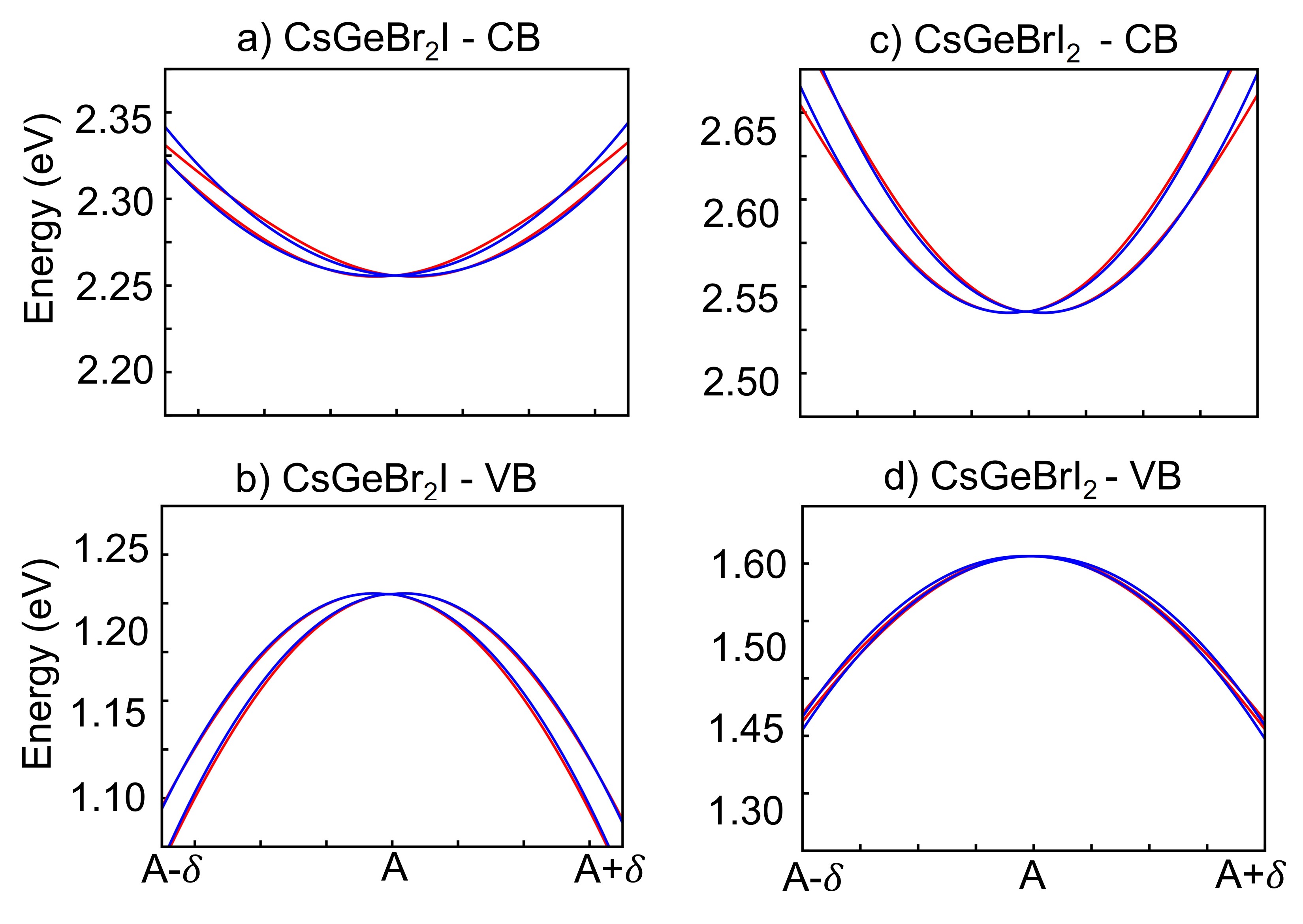}
    \caption{(a)-(b), (c)-(d) conduction and valence bands of CsGeBr$_2$I and CsGeBrI$_2$ for Cm phase respectively; Here red and blue curves are for DFT and \textbf{k.p} fitting parametrization around A (0, 0, 0.5) point.}
    \label{figS1}
\end{figure}

\begin{figure}
    \centering
    \includegraphics[width=1\linewidth]{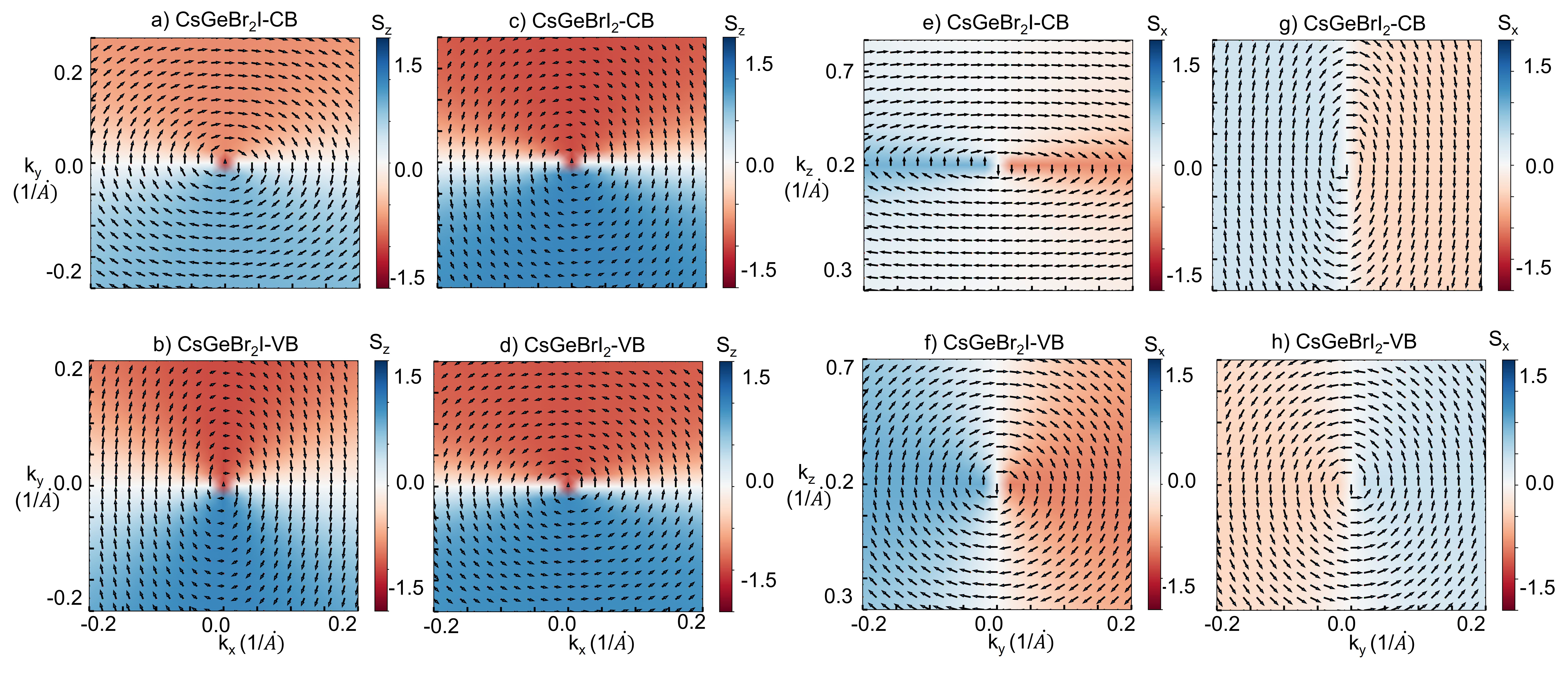}
    \caption{(a-b), (c-d) Spin textures obtained by k.p model Hamiltonian for conduction and valence bands of CsGeBr$_2$I and CsGeBrI$_2$ for Cm phase around A point along k$_x$-k$_y$ plane respectively. Similarly, (e-f) and (g-h) spin textures obtained by k.p model Hamiltonian for conduction and valence bands of CsGeBr$_2$I and CsGeBrI$_2$ for Cm phase around A point along k$_y$-k$_z$ plane respectively.}
    \label{figS2}
\end{figure}
\begin{figure}
    \centering
    \includegraphics[width=1\linewidth]{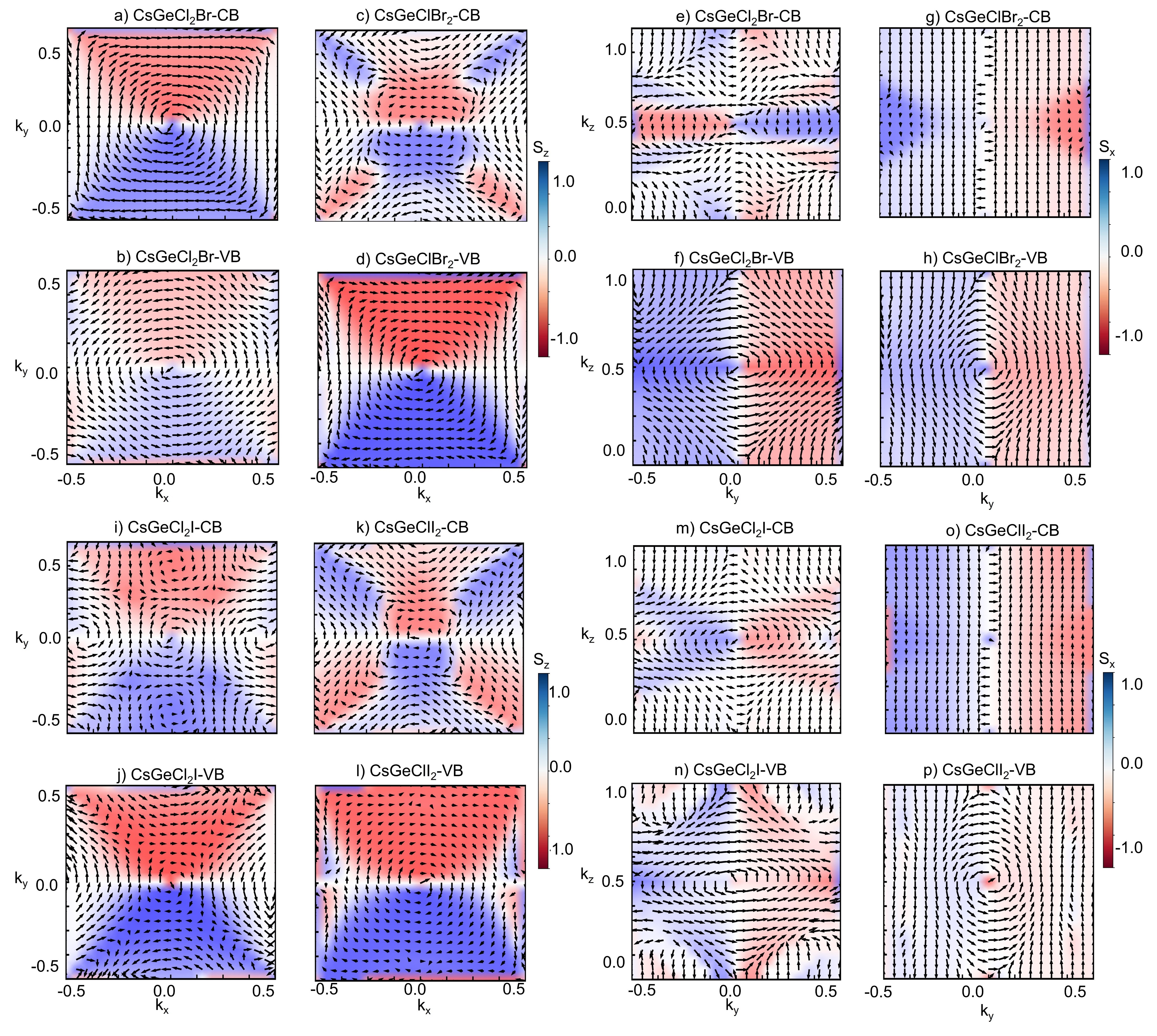}
    \caption{(a)-(p) DFT spin-textures of various chemically tuned structures for CB and VB in the k$_x$-k$_y$ and k$_y$-k$_z$ plane.}
    \label{figS3}
\end{figure}


\begin{thebibliography}{10}
\section*{References}
\bibitem{tai2019recent}
Q.~Tai, K.-C. Tang, and F.~Yan, ``Recent progress of inorganic perovskite solar cells,'' {\em Energy \& Environmental Science}, vol.~12, no.~8, pp.~2375--2405, 2019.

\bibitem{xiang2021review}
W.~Xiang, S.~F. Liu, and W.~Tress, ``A review on the stability of inorganic metal halide perovskites: challenges and opportunities for stable solar cells,'' {\em Energy \& Environmental Science}, vol.~14, no.~4, pp.~2090--2113, 2021.

\bibitem{han2024inorganic}
Q.~Han, J.~Wang, S.~Tian, S.~Hu, X.~Wu, R.~Bai, H.~Zhao, D.~W. Zhang, Q.~Sun, and L.~Ji, ``Inorganic perovskite-based active multifunctional integrated photonic devices,'' {\em Nature Communications}, vol.~15, no.~1, p.~1536, 2024.

\bibitem{chiara2021germanium}
R.~Chiara, M.~Morana, and L.~Malavasi, ``Germanium-based halide perovskites: materials, properties, and applications,'' {\em ChemPlusChem}, vol.~86, no.~6, pp.~879--888, 2021.

\bibitem{guo2024understanding}
Z.~Guo, M.~Yuan, G.~Chen, F.~Liu, R.~Lu, and W.-J. Yin, ``Understanding defects in perovskite solar cells through computation: current knowledge and future challenge,'' {\em Advanced Science}, vol.~11, no.~20, p.~2305799, 2024.

\bibitem{luo2019b}
B.~Luo, F.~Li, K.~Xu, Y.~Guo, Y.~Liu, Z.~Xia, and J.~Z. Zhang, ``B-site doped lead halide perovskites: synthesis, band engineering, photophysics, and light emission applications,'' {\em Journal of Materials Chemistry C}, vol.~7, no.~10, pp.~2781--2808, 2019.

\bibitem{yang2017spontaneous}
R.~X. Yang, J.~M. Skelton, E.~L. Da~Silva, J.~M. Frost, and A.~Walsh, ``{Spontaneous octahedral tilting in the cubic inorganic cesium halide perovskites CsSnX$_3$ and CsPbX$_3$ (X= F, Cl, Br, I)},'' {\em The journal of physical chemistry letters}, vol.~8, no.~19, pp.~4720--4726, 2017.

\bibitem{da2015phase}
E.~L. Da~Silva, J.~M. Skelton, S.~C. Parker, and A.~Walsh, ``{Phase stability and transformations in the halide perovskite CsSnI$_3$},'' {\em Physical Review B}, vol.~91, no.~14, p.~144107, 2015.

\bibitem{thiele1989kristallstrukturen}
G.~Thiele, H.~W. Rotter, and K.~Schmidt, ``{Die kristallstrukturen und phasentransformationen des tetramorphen RbGel$_3$},'' {\em Zeitschrift f{\"u}r anorganische und allgemeine Chemie}, vol.~571, no.~1, pp.~60--68, 1989.

\bibitem{yamada1994successive}
K.~Yamada, K.~Isobe, T.~Okuda, and Y.~Furukawa, ``Successive phase transitions and high ionic conductivity of trichlorogermanate (ii) salts as studied by 35c1 nqr and powder x-ray diffraction,'' {\em Zeitschrift f{\"u}r Naturforschung A}, vol.~49, no.~1-2, pp.~258--266, 1994.

\bibitem{thiele1987kristallstrukturen}
G.~Thiele, H.~W. Rotter, and K.~D. Schmidt, ``{Kristallstrukturen und phasentransformationen von caesiumtrihalogenogermanaten (II) CsGeX$_3$ (X= Cl, Br, I)},'' {\em Zeitschrift f{\"u}r anorganische und allgemeine Chemie}, vol.~545, no.~2, pp.~148--156, 1987.

\bibitem{kuang2025theoretical}
Z.~Kuang, B.~Liang, T.~Huang, T.~Shi, and W.~Xie, ``Theoretical advances and future perspectives of all-inorganic germanium-based perovskites,'' {\em Journal of Materials Chemistry C}, 2025.

\bibitem{yang2020assessment}
R.~X. Yang, J.~M. Skelton, E.~L. Da~Silva, J.~M. Frost, and A.~Walsh, ``Assessment of dynamic structural instabilities across 24 cubic inorganic halide perovskites,'' {\em The Journal of chemical physics}, vol.~152, no.~2, 2020.

\bibitem{zhang2022ferroelectricity}
Y.~Zhang, E.~Parsonnet, A.~Fernandez, S.~M. Griffin, H.~Huyan, C.-K. Lin, T.~Lei, J.~Jin, E.~S. Barnard, A.~Raja, {\em et~al.}, ``Ferroelectricity in a semiconducting all-inorganic halide perovskite,'' {\em Science advances}, vol.~8, no.~6, p.~eabj5881, 2022.

\bibitem{chen2022ferroelectric}
R.~Chen, C.~Liu, Y.~Chen, C.~Ye, S.~Chen, J.~Cheng, S.~Cao, S.~Wang, A.~Cui, Z.~Hu, {\em et~al.}, ``{Ferroelectric CsGeI$_3$ single crystals with a perovskite structure grown from aqueous solution},'' {\em The Journal of Physical Chemistry C}, vol.~127, no.~1, pp.~635--641, 2022.

\bibitem{kashikar2024coexistence}
R.~Kashikar, S.~Lisenkov, and I.~Ponomareva, ``{Coexistence of polar and antipolar phases in ferroelectric halide perovskite CsGeBr$_3$},'' {\em Physical Review B}, vol.~109, no.~2, p.~L020101, 2024.

\bibitem{kashikar2024ferroelectricity}
R.~Kashikar, A.~Valdespino, C.~Ogg, E.~Uppgard, S.~Lisenkov, and I.~Ponomareva, ``Ferroelectricity in ultrathin halide perovskites,'' {\em Nano Letters}, vol.~24, no.~34, pp.~10624--10630, 2024.

\bibitem{kashikar2026dft}
R.~Kashikar, J.~Townsend, S.~Lisenkov, and I.~Ponomareva, ``{DFT-based insight into finite-temperature properties of ferroelectric perovskites with lone-pair: the case of CsGeX$_3$ (X= Cl, Br, I)},'' {\em Journal of Physics: Condensed Matter}, vol.~38, no.~1, p.~015403, 2026.

\bibitem{CGX_Popoola}
A.~Popoola, N.~Maity, R.~Kashikar, S.~Lisenkov, and I.~Ponomareva, ``{Large Electrically and Chemically Tunable Rashba–Dresselhaus Effects in Ferroelectric CsGeX$_3$ (X = Cl, Br, I) Perovskites},'' {\em The Journal of Physical Chemistry C}, vol.~128, no.~41, pp.~17806--17812, 2024.

\bibitem{popoola2025mechanically}
A.~Popoola, R.~Kashikar, A.~Azmy, I.~Spanopoulos, H.~Jafari, J.~S{\l}awi{\'n}ska, S.~Witanachchi, S.~Lisenkov, and I.~Ponomareva, ``{Mechanically and electrically tunable Rashba-Edelstein effect in ferroelectric semiconductors, CsGeX$_3$ (X= I, Br, Cl)},'' {\em Physical Review Materials}, vol.~9, no.~8, p.~084412, 2025.

\bibitem{noheda2000stability}
B.~Noheda, D.~Cox, G.~Shirane, R.~Guo, B.~Jones, and L.~Cross, ``{Stability of the monoclinic phase in the ferroelectric perovskite PbZr$_{1-x}$Ti$_x$O$_3$},'' {\em Physical Review B}, vol.~63, no.~1, p.~014103, 2000.

\bibitem{noheda2001polarization}
B.~Noheda, D.~Cox, G.~Shirane, S.-E. Park, L.~Cross, and Z.~Zhong, ``{Polarization rotation via a monoclinic phase in the piezoelectric 92\% Pb${Zn}_{1/3}$Nb$_{2/3}$O$_3$-8\% PbTiO$_3$},'' {\em Physical Review Letters}, vol.~86, no.~17, p.~3891, 2001.

\bibitem{noheda2002structure}
B.~Noheda, ``Structure and high-piezoelectricity in lead oxide solid solutions,'' {\em Current Opinion in Solid State and Materials Science}, vol.~6, no.~1, pp.~27--34, 2002.

\bibitem{noheda2000tetragonal}
B.~Noheda, J.~Gonzalo, L.~Cross, R.~Guo, S.-E. Park, D.~Cox, and G.~Shirane, ``Tetragonal-to-monoclinic phase transition in a ferroelectric perovskite: The structure of pb$\mathrm{Zr}_{0.52}$$\mathrm{Ti}_{0.48}$o$_3$,'' {\em Physical Review B}, vol.~61, no.~13, p.~8687, 2000.

\bibitem{liu2020b}
M.~Liu, H.~Pasanen, H.~Ali-L{\"o}ytty, A.~Hiltunen, K.~Lahtonen, S.~Qudsia, J.-H. Sm{\aa}tt, M.~Valden, N.~V. Tkachenko, and P.~Vivo, ``B-site co-alloying with germanium improves the efficiency and stability of all-inorganic tin-based perovskite nanocrystal solar cells,'' {\em Angewandte Chemie International Edition}, vol.~59, no.~49, pp.~22117--22125, 2020.

\bibitem{azizman2024mixed}
M.~S.~A. Azizman, A.~W. Azhari, N.~Ibrahim, D.~S.~C. Halin, S.~Sepeai, N.~A. Ludin, M.~N.~M. Nor, and L.~N. Ho, ``Mixed cations tin-germanium perovskite: A promising approach for enhanced solar cell applications,'' {\em Heliyon}, vol.~10, no.~8, 2024.

\bibitem{raoui2021harnessing}
Y.~Raoui, S.~Kazim, Y.~Galagan, H.~Ez-Zahraouy, and S.~Ahmad, ``Harnessing the potential of lead-free sn--ge based perovskite solar cells by unlocking the recombination channels,'' {\em Sustainable Energy \& Fuels}, vol.~5, no.~18, pp.~4661--4667, 2021.

\bibitem{ito2018mixed}
N.~Ito, M.~A. Kamarudin, D.~Hirotani, Y.~Zhang, Q.~Shen, Y.~Ogomi, S.~Iikubo, T.~Minemoto, K.~Yoshino, and S.~Hayase, ``Mixed sn--ge perovskite for enhanced perovskite solar cell performance in air,'' {\em The journal of physical chemistry letters}, vol.~9, no.~7, pp.~1682--1688, 2018.

\bibitem{li2025enhanced}
A.~Li and W.~Liang, ``Enhanced charge separation and prolonged carrier lifetime in mixed sn--ge halide perovskite enabled by spontaneous symmetry breaking and moderate disorder,'' {\em Journal of Materials Chemistry A}, 2025.

\bibitem{townsend2024ferroelectric}
J.~Townsend, R.~Kashikar, S.~Lisenkov, and I.~Ponomareva, ``{Ferroelectric phases and phase transitions in CsGeBr$_3$ induced by mechanical load},'' {\em Physical Review B}, vol.~109, no.~9, p.~094121, 2024.

\bibitem{wang2024challenges}
Z.~Wang, Z.~Chen, R.~Xu, H.~Zhu, R.~Sundararaman, and J.~Shi, ``Challenges and opportunities in searching for rashba-dresselhaus materials for efficient spin-charge interconversion at room temperature,'' {\em Current Opinion in Solid State and Materials Science}, vol.~29, p.~101145, 2024.

\bibitem{bernevig2006exact}
B.~A. Bernevig, J.~Orenstein, and S.-C. Zhang, ``Exact su (2) symmetry and persistent spin helix in a spin-orbit coupled system,'' {\em Physical review letters}, vol.~97, no.~23, p.~236601, 2006.

\bibitem{schliemann2017colloquium}
J.~Schliemann, ``Colloquium: Persistent spin textures in semiconductor nanostructures,'' {\em Reviews of Modern Physics}, vol.~89, no.~1, p.~011001, 2017.

\bibitem{kohda2017physics}
M.~Kohda and G.~Salis, ``Physics and application of persistent spin helix state in semiconductor heterostructures,'' {\em Semiconductor Science and Technology}, vol.~32, no.~7, p.~073002, 2017.

\bibitem{PhysRevLett.90.146801}
J.~Schliemann, J.~C. Egues, and D.~Loss, ``Nonballistic spin-field-effect transistor,'' {\em Phys. Rev. Lett.}, vol.~90, p.~146801, Apr 2003.

\bibitem{Zunger1}
C.~M. Acosta, A.~Fazzio, G.~M. Dalpian, and A.~Zunger, ``Inverse design of compounds that have simultaneously ferroelectric and rashba cofunctionality,'' {\em Phys. Rev. B}, vol.~102, p.~144106, Oct 2020.

\bibitem{kashikar2023persistent}
R.~Kashikar, A.~Popoola, S.~Lisenkov, A.~Stroppa, and I.~Ponomareva, ``Persistent and quasipersistent spin textures in halide perovskites induced by uniaxial stress,'' {\em The Journal of Physical Chemistry Letters}, vol.~14, no.~38, pp.~8541--8547, 2023.

\bibitem{kresse1993ab}
G.~Kresse and J.~Hafner, ``Ab initio molecular dynamics for liquid metals,'' {\em Physical review B}, vol.~47, no.~1, p.~558, 1993.

\bibitem{kresse1996efficiency}
G.~Kresse and J.~Furthm{\"u}ller, ``Efficiency of ab-initio total energy calculations for metals and semiconductors using a plane-wave basis set,'' {\em Computational materials science}, vol.~6, no.~1, pp.~15--50, 1996.

\bibitem{kresse1996efficient}
G.~Kresse and J.~Furthm{\"u}ller, ``Efficient iterative schemes for ab initio total-energy calculations using a plane-wave basis set,'' {\em Physical review B}, vol.~54, no.~16, p.~11169, 1996.

\bibitem{kresse1999ultrasoft}
G.~Kresse and D.~Joubert, ``From ultrasoft pseudopotentials to the projector augmented-wave method,'' {\em Physical review b}, vol.~59, no.~3, p.~1758, 1999.

\bibitem{perdew1997generalized}
J.~P. Perdew, ``Generalized gradient approximation made simple,'' {\em Phys. Rev. Lett.}, vol.~77, p.~3868, 1997.

\bibitem{blochl1994projector}
P.~E. Bl{\"o}chl, ``Projector augmented-wave method,'' {\em Physical review B}, vol.~50, no.~24, p.~17953, 1994.

\bibitem{herath2020pyprocar}
U.~Herath, P.~Tavadze, X.~He, E.~Bousquet, S.~Singh, F.~Mu{\~n}oz, and A.~H. Romero, ``Pyprocar: A python library for electronic structure pre/post-processing,'' {\em Computer Physics Communications}, vol.~251, p.~107080, 2020.

\bibitem{togo2015first}
A.~Togo and I.~Tanaka, ``First principles phonon calculations in materials science,'' {\em Scripta materialia}, vol.~108, pp.~1--5, 2015.

\bibitem{togo2008first}
A.~Togo, F.~Oba, and I.~Tanaka, ``First-principles calculations of the ferroelastic transition between rutile-type and cacl$_2$-type sio$_2$ at high pressures,'' {\em Physical Review B—Condensed Matter and Materials Physics}, vol.~78, no.~13, p.~134106, 2008.

\bibitem{king1993theory}
R.~King-Smith and D.~Vanderbilt, ``Theory of polarization of crystalline solids,'' {\em Physical Review B}, vol.~47, no.~3, p.~1651, 1993.

\bibitem{stokes2005findsym}
H.~T. Stokes and D.~M. Hatch, ``Findsym: program for identifying the space-group symmetry of a crystal,'' {\em Applied Crystallography}, vol.~38, no.~1, pp.~237--238, 2005.

\bibitem{campbell2006isodisplace}
B.~J. Campbell, H.~T. Stokes, D.~E. Tanner, and D.~M. Hatch, ``Isodisplace: a web-based tool for exploring structural distortions,'' {\em Applied Crystallography}, vol.~39, no.~4, pp.~607--614, 2006.

\bibitem{momma2011vesta}
K.~Momma and F.~Izumi, ``Vesta 3 for three-dimensional visualization of crystal, volumetric and morphology data,'' {\em Applied Crystallography}, vol.~44, no.~6, pp.~1272--1276, 2011.

\bibitem{furness2020accurate}
J.~W. Furness, A.~D. Kaplan, J.~Ning, J.~P. Perdew, and J.~Sun, ``Accurate and numerically efficient r2scan meta-generalized gradient approximation,'' {\em The journal of physical chemistry letters}, vol.~11, no.~19, pp.~8208--8215, 2020.

\bibitem{kashikar2021chemically}
R.~Kashikar, P.~Ghosh, S.~Lisenkov, B.~Nanda, and I.~Ponomareva, ``Chemically and electrically tunable spin polarization in ferroelectric cd-based hybrid organic-inorganic perovskites,'' {\em Physical Review B}, vol.~104, no.~23, p.~235132, 2021.

\bibitem{mera2021different}
C.~Mera~Acosta, L.~Yuan, G.~M. Dalpian, and A.~Zunger, ``Different shapes of spin textures as a journey through the brillouin zone,'' {\em Physical Review B}, vol.~104, no.~10, p.~104408, 2021.

\end{thebibliography}
\end{document}